# Modulation of the Work Function by the Atomic Structure of Strong Organic Electron Acceptors on H-Si(111)

*Haiyuan Wang[1], Sergey V. Levchenko[3,1,4], Thorsten Schultz[2], Norbert Koch[2], Matthias Scheffler[1], and Mariana Rossi[1] \**

[1]Theory Department, Fritz-Haber-Institut der Max-Planck-Gesellschaft, 14195 Berlin, Germany
\*E-mail: rossi@fhi-berlin.mpg.de
[2]Institut für Physik & IRIS Adlershof, Humboldt-Universität zu Berlin, 12489 Berlin, Germany
[3]Center for Electrochemical Energy Storage, Skolkovo Institute of Science and Technology, 143026 Moscow, Russia
[4]Laboratory for Modeling and Development of New Materials, National University of Science and Technology ``MISIS'', 119049 Moscow, Russia

*Advances in hybrid organic/inorganic architectures for optoelectronics can be achieved by understanding how the atomic and electronic degrees of freedom cooperate or compete to yield the desired functional properties. Here we show how work-function changes are modulated by the structure of the organic components in model hybrid systems. We consider two cyano-quinodimethane derivatives (F4-TCNQ and F6-TCNNQ), which are strong electron-acceptor molecules, adsorbed on H-Si(111). From systematic structure searches employing range-separated hybrid HSE06 functional including many body van der Waals contributions, we predict that despite their similar composition, these molecules adsorb with significantly different densely-packed geometries in the first layer, due to strong intermolecular interaction. F6-TCNNQ shows a much stronger intralayer interaction (primarily due to van der Waals contributions) than F4-TCNQ in multilayered structures. The densely-packed geometries induce a large interface-charge rearrangement that result in a work-function increase of 1.11 and 1.76 eV for F4-TCNQ and F6-TCNNQ, respectively. Nuclear fluctuations at room temperature produce a wide distribution of work-function values, well modeled by a normal distribution with σ=0.17 eV. We corroborate our findings with*

*experimental evidence of pronounced island formation for F6-TCNNQ on H-Si(111 and with the agreement of trends between predicted and measured work-function changes.*

**Introduction** Silicon (Si) is the most commonly and widely used semiconductor device component in commercial microelectronics due to its high stability, abundance, and technological readiness. To achieve efficient device performance for optoelectronic and electronic applications such as solar cells[1] and field-effect transistors[2], appropriate energy level alignment and effective charge carrier transport across *p-n* junction structures are necessary. In particular, atomic arrangements and charge redistribution at interfaces are crucial design parameters for high-quality devices.[3] However, the design and performance of classical Si-based *p-n* junctions are limited in terms of length scale: Significant degradation of the device performance caused by short-channel effects, e.g. tunneling-induced leakage currents or drain-induced barrier lowering, can occur when approaching the nanoscale.[4]

To overcome such drawbacks, hybrid organic/inorganic systems can be a good alternative for Si-based nanoscale optoelectronic/electronic devices. Designing hybrid heterojunctions with specific electronic properties requires, primarily, the achievement of an appropriate energy level alignment of the comprising materials. In this respect, H-Si(111) surface, with its work function ($\Phi$) of about 4.3 eV,[5, 6] can be generally regarded as an anode material, when interfaced with charge acceptor layers having typically higher $\Phi$. To control the level alignment, one may add interlayers of molecular units. If these are composed by strong electron acceptors, there will be electron transfer to the molecular layer and hole accumulation in the Si anode layer. Good models for the organic electron acceptor component are 2,3,5,6-tetrafluoro-7,7,8,8-tetracyano-quinodi-methane (F4-TCNQ) and 1,3,4,5,7,8-hexafluorotetracyano-naphthoquinodimethane (F6-TCNNQ) due to their high electron affinity of 5.24 eV[7] and 5.60 eV[8], respectively, as measured in neat molecular thin films.

The quantitative tuning of the properties of such interfaces depends critically on the structure of the organic/inorganic interface, which is an aspect that is rarely properly modelled or measured under realistic device conditions.[9-12] In this contribution, we report an experimental and theoretical analysis of the interface between F4-TCNQ/F6-TCNNQ and the well-defined H-Si(111) surface, with the purpose of understanding the impact of coverage and different geometrical arrangements of the organic layers on the electronic structure of the interface, including effects of temperature.

On the theory side, we perform calculations within the all-electron, full-potential, numeric atom-centered basis framework of the FHI-aims code.[13] We employ density-functional theory (DFT) using the screened hybrid Heyd-Scuseria-Ernzerhof (HSE06) functional,[14] augmented with many-body dispersion (MBD) van der Waals interactions.[15] This functional significantly improves the description of the Kohn-Sham single-particle energy levels compared with semilocal approximations, due to the mitigation of the charge delocalization error.[16] In particular, for the systems considered here, the Perdew-Burke-Ernzerhof (PBE) generalized gradient approximation[17] underestimates the band gap, resulting in an unphysically large charge transfer to the organic layer, as schematically shown in **Figure** 1. Experimentally, we characterize these interfaces with X-ray (XPS), ultraviolet photoelectron spectroscopy (UPS) and scanning force microscopy (SFM).

In the following, we show the characterization of the atomic structure and the related electronic properties of F4-TCNQ and F6-TCNNQ on H-Si(111), highlighting the similarities and differences of the theoretical models and the experiment setup. We show how and why they exhibit different adsorption geometries and unravel the physical reasons that lead to different amounts of charge transfer and interface dipoles that modulate the work function.

***Theoretical Results*** In our theoretical models, we consider the interface of F4-TCNQ and F6-TCNNQ with pristine, undoped H-Si(111). Since the atomic structure and the electronic properties are interconnected, we performed a systematic structure search employing a grid search algorithm. The search involved 11 types of unit cells (containing 1 and 2 molecules per cell) and 124 structures in total (79 for F4-TCNQ and 45 for F6-TCNNQ). Details of this search are reported in Figures S1 and S2 in Supporting Information (SI). As shown in **Figure 2**a, the interfaces of F4-TCNQ or F6-TCNNQ with H-Si(111) exhibit significantly different structural arrangements under distinct molecular packing densities of a self-assembled monolayer. We define the packing density $\theta_p$ as the number of molecules in the unit cell divided by the number of atoms in the top Si layer in the respective cell. We regard only interaction energies, assuming a fixed amount of molecules per unit area and disregarding an equilibrium with an external reservoir.

For both molecules, at low $\theta_p$ (below approximately 11%), the energetically most stable adsorption configuration is a flat-lying geometry in which the symmetric planes of the molecules tend to follow hydrogen rows, and N atoms prefer to be in the center of the triangle enclosed by three surface hydrogens shown in Figure S3 in SI. We also observe that F4-TCNQ (F6-TCNNQ) undergoes a pronounced geometrical distortion in the $z$-direction (normal to the interface), resulting in a vertical bending, $\Delta d$, of approximately 0.4 Å (0.5 Å) due to interaction with the surface.

As shown in Figure 2a, at $\theta_p$ between about 15% and 25%, the energetically most stable structures of F4-TCNQ change from lying-flat to significantly inclined, with the short axis of the molecule forming an acute angle to the surface normal ("short-tilted"). At $\theta_p \geq 25\%$, the most stable geometries are tilted molecules with the long axis forming an acute angle to the surface normal ("long-tilted"). Within the models studied here, the most stable structure over all packing densities is found at $\theta_p = 25\%$ where the long axis of the molecule presents an angle $\beta \approx 63°$ to the surface normal. For F6-TCNNQ, on the other hand, while an increase in packing density also stabilizes geometries where molecules stand on the surface, the most stable conformation is found to be the one where the short axis of the molecule forms an angle $\beta \approx 35°$ to the surface normal, also at $\theta_p = 25\%$. The phases at much higher packing densities, e.g. 50%, become less stable, primarily due to steric effects.

To understand the physical reasons behind the preferred adsorption geometries, the most stable structures at each packing density can be regarded as the result of the energy balance between the molecule-substrate (labeled by green diamonds in Figure 2b,c) and molecule-molecule (labeled by blue triangles in Figure 2b,c) interactions. Figures 2b and 2c show that the interaction between molecules and substrate dominates at low $\theta_p$, whereas at $\theta_p \approx 20\%$ the molecule-molecule interaction starts to dominate. We also found that the interaction between molecules is slightly repulsive when molecules lie down on the substrate, whereas it is strongly attractive at tilted configurations. This behavior of molecule-molecule interaction can be understood by the net charge of the molecules, as discussed further below.

The fact that these molecular acceptors strongly prefer this tightly packed arrangement implies that they may tend to form islands on this surface. We thus investigated whether multilayers or more tightly packed geometries (with random relative angles) within the first layer are energetically favored. For the tighter packing in the first layer, a $2\sqrt{3} \times \sqrt{3}$ H-Si(111) unit cell with two molecules in the unit cell ($\theta_p \approx 33.3\%$) was considered. For modeling multilayers, we considered the most stable monolayer structure and added additional layers in different orientations (models shown in **Figure** 3a). The adsorption energies per layer for these models are illustrated in Figure 3b. The results reveal that forming the second layer is energetically more favorable than the first layer by 0.2 eV/molecule for F4-TCNQ, but less stable than the monolayer ordered structure for F6-TCNNQ. The molecular phases for the tighter packing of the first layer are found to be considerably less energetically favored. Another important point is that additional layers of F4-TCNQ/H-Si(111) show no preference to be ordered in parallel or zigzag stacking arrangements. For F6-TCNNQ/H-Si(111), instead, the second layer prefers to be oriented in a zigzag arrangement, while at the fourth layer the parallel arrangement dominates.

The layer formation energy can also be decomposed in an interlayer and intralayer component, as shown in Figure 3b. This decomposition shows that while for F4-TCNQ both interlayer and intralayer interactions are comparable up to 4 layers, in the case of F6-TCNNQ the intralayer interaction is 0.6 to 0.7 eV stronger than the interlayer interaction for the second and third layers. Additionally, we find that this difference in the intralayer interaction between the two cases is predominantly due to intermolecular van der Waals (vdW) interactions. The intralayer contribution from vdW is twice stronger for F6-TCNNQ than for F4-TCNQ, while the interlayer contribution is similar, as shown in Figure 3c.

Having unraveled the atomic structure of these interfaces, we focus on their electronic properties. The charge rearrangement of the heterosystem, $\Delta\rho$, can be obtained by calculating $\Delta\rho = \rho_{full} - \rho_{surf} - \rho_{mol}$ where $\rho$ represents the electron density of each system (full, clean surface, and molecular layer) integrated in the directions parallel to the surface. This quantity at low and high packing densities for F6-TCNNQ/H-Si(111) is shown in **Figure 4**a,b. Positive and negative values indicate electron accumulation and depletion, respectively. This analysis demonstrates that electron density is transferred from the surface Si and H atoms to the molecules. Similar results for F4-TCNQ/H-Si(111) are presented in Figure S5a,b in SI. The charge on a molecule is obtained by integrating $\Delta\rho$ and finding the point of charge compensation at the interface. We predict F4-TCNQ (F6-TCNNQ) to be charged with 0.36 (0.40) electrons when lying flat on the surface (low-coverages) and with 0.07 (0.12) electrons at the higher packing densities. The charge density per area remains approximately constant for F4-TCNQ (around $1.4 \times 10^{-3}$ e/Å$^2$) and increases slightly at high packing densities for F6-TCNNQ ($1.6 \times 10^{-3}$ e/Å$^2$ at the $\theta_p$ of 5% and $2.4 \times 10^{-3}$ e/Å$^2$ at 25%).

To identify the electronic levels that play a key role at the interface, the projected density of states (PDOS) for F6-TCNNQ/H-Si(111) is shown in Figure 4c. The top and bottom panels stand for the high and low packing densities. Through a comparison to the electronic levels in the isolated molecule and the expected level alignment, we can identify the orbitals that corresponded to the lowest unoccupied molecular orbital (LUMO), the highest occupied molecular orbital (HOMO), and the HOMO-1 before contact. It reveals that the initially empty LUMO in the isolated molecule becomes partially filled upon adsorption on H-Si(111) surface due to the electron transfer from the Si substrate. For low $\theta_p$, we also observe a symmetry breaking of the spin-up and spin-down channels, which is not manifested at higher $\theta_p$. With respect to the electronic properties, F4-TCNQ behaves in a similar way even if the structures at high $\theta_p$ are different (Figure S5c in SI). The PDOS for multilayered F6-TCNNQ on H-Si(111), shown in Figure 4d, corroborates that the (small) charge transfer exclusively takes place within the first molecular layer, while the other layers of F6-TCNNQ remain neutral. The shift of the molecular PDOS of the second and third layers to higher energies serves as an effective probe of the dipole induced by the first layer.

We calculated the N1$s$ core level shifts (CLS) between the lying-down ($\theta_p$ = 5%) and tilted ($\theta_p$ = 25%) configurations (case 1), and also between the lying-down molecule and the zigzag configuration of multilayers adsorbed on H-Si(111) (case 2) using the Slater-Janak transition state approximation[19] (details in Table S1 in the SI). We considered N atoms in the first molecular layer and CLSs are calculated to be 0.68 eV and 0.95 eV for F4-TCNQ/H-Si(111), 1.14 eV and 1.41 eV for F6-TCNNQ/H-Si(111) in case 1 and case 2, respectively.

On the basis of the charge redistribution at the interface between the H-Si(111) and molecules, we calculated the $\Delta\Phi$ as functions of $\theta_p$ and the multilayer models for each adsorbed molecule, as illustrated in Figure 4e. The calculated results show that the $\Delta\Phi$ increases with two molecular layers (ML), and saturates with more layers (> 2MLs) for both molecules. We predict a work function increase of 1.11 eV for F4-TCNQ and 1.76 eV for F6-TCNNQ. The calculated $\Delta\Phi$ can be decomposed into a contribution of the molecular dipole (MD) and the interface charge rearrangement (CR) ($\Delta\Phi = \Delta\Phi_{MD} + \Delta\Phi_{CR}$).[9,10] As reported in **Table 1**, only for the case of low $\theta_p$ ($\theta_p$ = 5%), $\Delta\Phi_{MD}$ has a non-zero contribution due to the bending of the molecule that produces a small positive dipole moment in the $z$-axis (0.17 eÅ for F4-TCNQ and 0.20 eÅ for F6-TCNNQ). This leads to a $\Phi$ decrease of 0.12 eV (0.15 eV) for the F4-TCNQ (F6-TCNNQ)/H-Si interface. On the other hand, the charge rearrangement at the interface induces a negative dipole moment in the *z*-direction (pointing towards the surface), resulting in band-bending and a $\Phi$ increase of 0.83 eV (0.91 eV) for F4-TCNQ (F6-TCNNQ). For high $\theta_p$ of one molecular layer ($\theta_p = 25\%$), only $\Delta\Phi_{CR}$ plays a dominant role, since there is no permanent dipole on the molecules. The presence and absence of this molecular dipole is the main reason for the increase in $\Phi$ when going from $\theta_p$ of 5% to 25%. In F6-TCNNQ/H-Si(111) not only the absence of the molecular dipole, but also the increased charge transfer to the organic layer contribute to the larger $\Delta\Phi$ at $\theta_p = 25\%$.

To examine the impact of temperature on $\Delta\Phi$, we show the distribution of $\Phi$ values obtained from *ab initio* molecular dynamics simulations at 300 K for systems containing one and two ML of F4-TCNQ adsorbed on 4 bilayers of the H-Si(111) surface as illustrated in Figure 4f. These simulations were performed with the PBE+MBD functional, and the calculated $\Delta\Phi$ values were shifted to reflect HSE06+MBD values (see Methods). Temperature indeed affects the average value of $\Delta\Phi$ for the adsorbed monolayer, causing an increase in $\Delta\Phi$ by approximately 0.13 eV. For the bilayer, the shift is slightly smaller, of approximately 0.07 eV. Even more striking, however, is that the variation of the overall $\Phi$ over a time window encompassing a few ps is almost 1 eV. By fitting a Gaussian to these distributions, we determined the standard deviation to be $\sigma = 0.17$ eV at room temperature, corresponding to 6.5 times the thermal energy.

*Experimental Results* Experiments were conducted on a *n*-doped H-Si(111) substrate, with doping concentration of $10^{13}$-$10^{14}$ cm$^{-3}$, with 4% C and 2% O residuals in the substrate used for F4-TCNQ deposition and 10% C and 7% O residuals in the substrate used for F6-TCNNQ deposition (XPS), as well as step terraces of 2.7Å height (SFM). The work function of the bare H-Si(111) was determined to be 4.3±0.2 eV by the extrapolation of the secondary electron cut-off (SECO) obtained by UPS, consistent with values reported in literature.[S1, S2]

The binding energy of the Si2p$_{3/2}$ peak was 99.1±0.1 eV, indicating an initial upward band bending by approximately 0.4±0.1 eV.[S2] This surface band bending is caused by the presence of surface states, which are attributed to unsaturated dangling bonds.[S3] After molecular adsorption a further upward band bending of approximately 0.2 eV was observed (see SI). SFM images of H-Si(111) covered with a *nominal* thickness (calculated from the deposited mass and assuming layer by layer growth) of 1.2 nm of F4-TCNQ and 1.6 nm of F6-TCNNQ are shown in **Figure 5**. It is observed that F4-TCNQ forms plateaus consistent with the existence of areas containing 1 to 3 layers of molecules, but F6-TCNNQ forms pronounced tall islands. The average height of these islands is 40 nm and they cover approximately 3 % of the surface.

The XPS N1s core level spectra after molecule adsorption exhibit two distinct peaks, which are assigned to neutral and charged molecules at higher and lower binding energy, respectively, as shown in **Figure** 6a. The energy difference between the centers of these peaks is around 1.4 eV for F4-TCNQ and 1.7 eV for F6-TCNNQ. From the area ratio of the two peaks, the fraction of charged molecules is estimated to be approximately 38% for F4-TCNQ and 29% for F6-TCNNQ. It should be noted that most of the signal derives from the monolayer (or few layers) areas.

The ΔΦ of both F4-TCNQ/H-Si(111) and F6-TCNNQ/H-Si(111) are shown in Figure 6b. The values of Φ at different nominal coverages were obtained through the secondary electron cutoff (SECO) by UPS (details in Figure S11 in SI) and the initial measured band bending of the bare substrate was considered as an offset to calculate ΔΦ. The results show that ΔΦ saturates gradually as a function of nominal coverage and this saturation is slower for F6-TCNNQ than for F4-TCNQ, indicating different growth modes.

*Discussion* Despite the differences in the experimental setup and the theoretical models that we consider, there are qualitative trends that can be compared between these sets of data and that provide a deep understanding of the physics in this hybrid organic/inorganic setup. The strong island formation propensity observed in the AFM measurements of F6-TCNNQ and the smoother multilayer coverage of F4-TCNQ can be connected to the much stronger intralayer interaction of F6-TCNNQ, most of which is attributed to van der Waals interactions. Even without simulating the kinetics of island formation, this strong intermolecular enthalpic contribution in F6-TCNNQ is most likely the main contribution that drives molecular aggregation and wins over a smoother multilayer formation, as in the case of F4-TCNQ. The fact that we only consider relatively small unit-cell areas in this study means that the exact atomic arrangement of the molecular layers may be slightly different in reality. However, we expect the trends in structural arrangements at different packing densities to hold even in more complex architectures.

Regarding the orientation of the adsorbates, in Ref. 12, the structure of F4-TCNQ/H-Si(111) was characterized by means of angle-dependent infrared spectroscopy. In that case, F4-TCNQ was deposited on H-Si(111) by the wet chemical method, such that the first layer is more consistent with the low packing-densities in our theoretical model. They conclude that at those packing densities, the first layer is composed of molecules lying flat on the substrate (consistent with our results), and that multilayer structures are not oriented in any particular order, in agreement with our calculations.

Regarding the electronic structure, the experimental N1s CLS suggest that neutral and negatively charged molecules coexist within the first layers. A comparison with calculations suggest two possible scenarios. One scenario would be that there are areas of the surface where the organic molecules are adsorbing in relatively isolated flat-lying configurations and thus more negatively charged, and others where they adsorb tilted and within multilayer islands or plateaus – in which the bottom layer is charged but the higher layers are not. This inhomogeneity could stem from steps and defects on the surface and cannot be ruled out for the samples investigated here. The other scenario would be, instead, that the electrons transferred from the surface to the adsorbed molecules distribute unevenly between the molecules. Such disproportionation has been reported previously for other systems.[20] We considered up to three tilted molecules in the unit cell and included larger fractions of exact exchange in the HSE06 functional (up to 100%), but no charge disproportionation was found. This indicates that the first scenario may be correct, or there is a fundamental difference between the charge transfer behavior between the doped and undoped substrate. Due to the high density of states close to the VBM in this system we expect this not to be the case, but cannot rule it out. We conclude that in order to fully resolve this issue, complementary experiments with varying surface defect density and molecular coverage, as well as a broader large-cell combinatorial structure search including defects and doping would be needed, which are beyond the scope of this manuscript. Nonetheless, both scenarios are compatible with all other analysis and conclusions in this work.

The calculated values for $\Delta\Phi$ on pristine H-Si(111) can only be qualitatively compared to the experimental $\Delta\Phi$, where steps, defects, and impurities (as well as inherent *n*-doping) are known to be present. It is nevertheless encouraging that the calculated values are quite close to experiment and we find agreement in the trend that F6-TCNNQ induces a larger $\Delta\Phi$ than F4-TCNQ. Additionally, our theoretical models are consistent with the SK island growth mode of F6-TCNNQ and we can observe that temperature shifts $\Delta\Phi$ upwards for F4-TCNQ. We stress that using the HSE06(+MBD) functional (instead of, e.g., PBE) and considering multiple layers is essential to obtain even this qualitative agreement. We observe that the large variation of the interface dipole due to nuclear fluctuations at finite temperatures can impact the band offset and level-alignment in semiconductor interfaces. For the design of electronic devices based on organic/inorganic hybrid materials, this temperature effect is a crucial parameter to consider, and experimental reports of unusual temperature dependences of Schottky-barrier heights that could be related to such broad work-function distributions have been recently reported.[21]

For setting our results into a broader context, we compare experimental ΔΦ values induced by F4-TCNQ and F6-TCNNQ deposited on a wide range of different substrates, as reported in the literature, in Figure 6c. This comparison shows a clear correlation between the initial work function $\Phi_0$ of the substrate and the eventual ΔΦ, namely, a higher ΔΦ can be achieved for lower $\Phi_0$. More interestingly, it is possible to identify two different slopes in this correlation, i.e., there is a steeper slope for semiconductors than that for metals. We assign this difference, tentatively, to the fact that the (change of) surface band-bending at interfaces with semiconductors contributes to the final ΔΦ.[22] Additionally, the weak dependence of ΔΦ to the $\Phi_0$ of metallic substrate points to the existence of Fermi-level pinning. Our H-Si(111) substrate interfaced with F4-TCNQ and F6-TCNNQ exhibit a large ΔΦ (around 1.6 eV), which is slightly smaller than the ones achieved for polar ZnO substrates, as ZnO can feature much larger induced surface band bending. However, inevitable defect formation, which induces chemisorption and property-instability, significantly limits the practical usage of ZnO substrates.[23] In contrast, H-Si(111) does not exhibit pronounced thermal and chemical degradation, having a high degree of stability in ambient atmosphere, deionized water or even highly acidic or highly basic solutions.[24] Hence, hybrid interfaces based on H-Si(111) are good candidates for further research and the development of advanced optoelectronic/electronic devices.

***Conclusions*** In this work, we have demonstrated the adsorption behavior of F4-TCNQ and F6-TCNNQ on H-Si(111) at different packing densities and their associated interplay with the electronic structure and charge rearrangements of the interfaces. Although it is widely assumed that F4-TCNQ and F6-TCNNQ show similar characteristics at these interfaces, our present work identifies significant differences in the behavior of these molecules with respect to molecular configuration, island growth characteristics, and the degree of electronic reconstruction. We show from theory that F4-TCNQ prefers a long-tilted conformation and F6-TCNNQ adopts a short-tilted conformation on the first layer interacting with pristine H-Si(111). The densely packed structures are favored by molecule-molecule interactions, which dominate at high packing densities. For multilayer structures, the intralayer interactions are much stronger in F6-TCNNQ than F4-TCNQ, while the interlayer interactions are comparable. In the case of F6-TCNNQ, 80% of the intralayer interaction stems from vdW interactions, which are twice stronger than for F4-TCNQ. The calculated $\Delta\Phi$ at higher packing densities is mainly due to the absence of any molecular dipole induced by deformation and the we find that the second layer still contributes to $\Delta\Phi$. For F6-TCNNQ we also observe an increased charge transfer per area at high packing densities. Temperature is shown to further increase $\Delta\Phi$ by a couple of hundreds of meV, which is stems from interplay of the electronic structure with the nuclear vibrations in the rather flexible molecular layers.

The theoretical predictions enlighten different aspects of measurements that were conducted in these interfaces. In particular, SFM measurements confirm the existence of tall islands consistent with SK island growth for F6-TCNNQ. For F4-TCNQ, only smoother multilayer structures were observed. Our theoretical results suggest that the much stronger intermolecular intralayer interaction of F6-TCNNQ (if compared to F4-TCNQ) in densely-packed structures can explain this different behavior. Despite the complexity of these interfaces in experiment, the absolute values of $\Delta\Phi$ calculated for pristine interfaces compare well to the measured ones. The higher $\Delta\Phi$ in the F6-TCNNQ interface observed experimentally can be related to the calculated larger amount of charge transfer per area in this interface, if compared to F4-TCNQ.

The findings presented here compose a very comprehensive assessment of the nuclear and electronic structure of these interfaces, providing further evidence that the nuclear structure can be almost as important as the chemical composition to define the electronic properties of such systems. In particular, such studies involving multilayer organic structures and the decomposition of the interaction energy in different components can identify the physical effects that induce molecular aggregation. This paves the way to a better control and design of highly efficient organic/inorganic hybrid systems for device applications.

**Methods**

*Theoretical methods:* All calculations presented were based on DFT.[25] The PBE generalized gradient approximation[26] functional was used for structure pre-relaxation, and the hybrid HSE06[14] functional was used for the post-relaxation and electronic properties, within the framework of the all-electron, numeric atom-centered electronic-structure package FHI-aims[13]. For the PBE calculations, "tight" basis sets and numerical settings were used, while for HSE06 we used the "LVL-intermediate" settings in the FHI-aims package. All calculations include many-body dispersion interactions (MBD).[15] The dipole correction was employed to prevent the artificial electrostatic interactions between the periodic supercells.[27] A 10 × 10 × 1 *k*-point sampling was used for the (1 × 1) H-Si(111) slab and scaled appropriately for larger supercells to guarantee consistent convergence. Five double-layers of Si(111) surface with H passivated on the top and bottom Si layers were considered for our model slabs, and a vacuum width of 100 Å was used. We emphasize that especially for multilayer systems, our setup with a large vacuum region was necessary. We checked that for the interfaces we are considering, all electrostatic effects are included in the simulation cells, due to the high DOS around the VBM (see SI). Two double layers of Si at the bottom were fixed, while the rest of Si layers, as well as the two H layers, were allowed to fully relax. The adsorption energy is defined as $E_a = (E_{full} - E_{surf} - N_{mol}E_{mol})/N_{mol}$ and the energy difference for multilayer structures is calculated as $\Delta E = E^{i+1} - E^i - E_{mol}$ where $E^i$ is the total energy of the interface containing *i* organic layers and $E_{mol}$ is the total energy of the isolated molecule.

*Ab initio* molecular dynamics simulations were performed using the interface between i-PI[28] and FHI-aims, in the NVT ensemble using a stochastic velocity-rescaling thermostat[29] and a 0.5 fs time-step. In these simulations we considered 8 Si layers in the slab and fixed only the bottom two. We thermalized the systems for 2 ps and gathered statistics from subsequent 10 ps of simulations in each case. For the zigzag configuration of 2-layers of F4-TCNQ adsorbed on H-Si(111), we have chosen a few conformations from our PBE+MBD simulations and calculated the Φ with the HSE06+MBD functional. The correlation between these two values is shown in the **Figure S7**a in SI. The relationship is very close to linear over a wide range of values, and represents a shift of 0.11 eV. From our calculations, we also know that for the clean H-Si(111) surface, changing from PBE to HSE06 corresponds to a Φ increase of about 0.3 eV. We have therefore rigidly shifted our distributions to reflect these ΔΦ values with the HSE06. The structure changes related with temperature were also plotted in Figure S7b in SI.

*Sample preparation and PES measurements*: All the experiments were performed at Humboldt-Universität zu Berlin, Germany. The Si(111) samples ($N_D 10^{13}$-$10^{14}$ cm$^{-3}$) were sonicated in isopropanol and acetone for 10 min each, before they were immersed in a 40 % $NH_4F$ solution for 800 s, to remove the surface oxide layer and saturate the remaining dangling bonds with hydrogen. Afterwards the samples were transferred to an UHV system (base pressure $10^{-9}$ mbar) as fast as possible (air exposure <5 min). They were annealed at 400°C to remove remaining surface contaminations. Most photoelectron spectroscopy characterizations were done using an Omicron 75 hemispherical energy analyzer with an energy resolution of 150 meV. As excitation sources, the K radiation (1253.69 eV) of a magnesium anode was used for XPS and the He I radiation (21.21 eV) of a Helium-discharge lamp was used for UPS. During SECO measurements a voltage of -10 V was applied between sample and analyzer. Additional high resolution XPS spectra were recorded at a JEOL JPS-9030, employing monochromatized Al K radiation (1486.58 eV). The molecules were evaporated from resistively heated quartz crucibles and their nominal thickness was determined from a quartz-crystal microbalance.

*Scanning force microscopy*: The SFM measurements were performed with a Bruker Dimension Icon with Scan-Asyst. All measurements were performed in air, employing PeakForce Tapping mode and ScanAsyst-Air cantilevers. The data were evaluated with Gwyddion.[30]

**Supporting Information**
Supporting Information is available from the Wiley Online Library or from the author.

All of calculations can be found in the NOMAD repository:
http://dx.doi.org/10.17172/NOMAD/2018.10.05-1

**Acknowledgments**
The authors acknowledge financial support from the SFB-951 (HIOS) project of the Deutsche Forshungsgemeinschaft.

**References**

[1] a) M. A. Green, A. W. Blakers, J. Q. Shi, E. M. Keller, S. R. Wenham, *Appl. Phys. Lett.* **1984**, *44*, 12; b) A. W. Blakers, A. H. Wang, A. M. Milne, J. H. Zhao, M. A. Green, *Appl. Phys. Lett.* **1989**, *55*, 1363; c) J. H. Oh, H. C. Yuan, H. M. Branz, *Nat. Nanotechnol.* **2012**, *7*, 743.

[2] a) W. E. Spear, P. G. LeComber, *J. Non-Cryst. Solids* **1972**, *8*, 727; b) R. Weis, B. Müller, P. Fromherz, *Phys. Rev. Lett.* **1996**, *76*, 327; c) J. Xiang, W. Lu, Y. J. Hu, Y. Wu, H. Yan, C. M. Lieber, *Nature* **2006**, *441*, 489.

[3] a) B. O'regan, M. Grätzel, *Nature* **1991**, *353*, 737; b) H. Sirringhaus, *Adv. Mater.* **2005**, *17*, 2411.

[4] a) P. S. Peercy, *Nature* **2000**, *406*, 1023; b) R. Chau, B. Doyle, M. Doczy, S. Datta, S. Hareland, *Device Research Conf.* **2003**, *123*, 2003; c) P. H. -S. Wong, *Solid-State Electron.* **2005**, *49*, 755; d) D. Wang, B. Sheriff, J. R. Heath, *Nano Lett.* **2006**, *6*, 1096.

[5] a) R. Hunger, Chr. Pettenkofer, R. Scheer, *J. Appl. Phys*. **2002**, *91*, 6560; b) R. S. Becker, G. S. Higashi, Y. J. Chabal, A. J. Becker, *Phys. Rev. Lett.* **1990**, *65*, 1917.

[6] D. C. Gleason-Rohrer, B. S. Brunschwig, N. S. Lewis, *J. Phys. Chem. C* **2013**, *117*, 18031.

[7] a) W. Gao, A. Kahn, *J. Appl. Phys.* **2003**, *94*, 359; b) W. Gao, A. Kahn, *Appl. Phys. Lett.* **2001**, *79*, 4040.

[8] F. Y. Zhang, A. Kahn, *Adv. Funct. Mater.*, **2018**, *28*, 1703780.

[9] L. Romaner, G. Heimel, J. L. Brédas, A. Gerlach, F. Schreiber, R. L. Johnson, J. Zegenhagen, S. Duhm, N. Koch, E. Zojer, *Phys. Rev. Lett.* **2007**, *99*, 256801.

[10] G. M. Rangger, O. T. Hofmann, L. Romaner, G. Heimel, B. Bröker, R. P. Blum, R. L. Johnson, N. Koch, E. Zojer, *Phys. Rev. B* **2009**, *79*, 165306.

[11] C. Christodoulou, A. Giannakopoulos, G. Ligorio, M. Oehzelt, M. Timpel, J. Niederhausen, L. Pasquali, A. Giglia, K. Parvez, K. Müllen, D. Beljonne, N. Koch, M. V. Nardi, *ACS Appl. Mater. Interfaces* **2015**, *7*, 19134.

[12] M. Furuhashi, J. Yoshinobu, *J. Phys. Chem. Lett.* **2010**, *1*, 1655.

[13] a) V. Blum, R. Gehrke, F. Hanke, P. Havu, V. Havu, X. Ren, K. Reuter, M. Scheffler, *Comput. Phys. Commun.* **2009**, *180*, 2175. b) S. V. Levchenko, X. G. Ren, J. Wieferink, R. Johanni, P. Rinke, V. Blum, M. Scheffler, *Comput. Phys. Commun.* **2015**, *192*, 60.

[14] A. V. Krukau, O. A. Uydrov, A. F. Izmaylov, and G. E. Scuseria, *J. Chem, Phys.* **2006**, *125*, 224106.

[15] a) A. Tkatchenko, R. A. DiStasio, Jr., R. Car, M. Scheffler, *Phys. Rev. Lett.* **2012**, *108*, 236402; b) A. Tkatchenko, A. Ambrosetti, R. A. DiStasio Jr. *J. Chem. Phys.* **2013**, *138*, 074106.

[16] P. Mori-Sánchez, A. J. Cohen, W. T. Yang, *Phys. Rev. Lett.* **2008**, *100*, 146401.

[17] a) J. P. Perdew, M. Levy, *Phys. Rev. Lett.* **1983**, *51*, 1884; b) L. J. Sham M. Schlüter, *Phys. Rev. Lett.* **1983**, *51*, 1888.

[18] D. Bimberg, R. Blachnik, M. Cardona, P.J. Dean, T. Grave, G. Harbeke, K. Hübner, U. Kaufmann, W. Kress, O. Madelung, W. von Münch, U. Rössler, J. Schneider, M. Schulz, M.S. Skolnick, *Physics of Group IV Elements and III-V Compounds*, Vol. 17 (Eds: K. -H. Hellwege, O. Madelung), Springer-Verlag Berlin Heidelberg **1982**, p. 17a.

[19] a) J. F. Janak, *Phys. Rev. B* **1978**, *18*, 7165; S. Lizzit, A. Baraldi, A. Groso, K. Reuter, M. V Ganduglia-Pirovano, C. Stampfl, M. Scheffler, M. Stichler, C. Keller, W. Wurth, D. Menzel, *Phys. Rev. B* **2001**, *63*, 205419.

[20] a) O. T. Hofmann. P. Rinke, M. Scheffler, G. Heimel, *ACS Nano* **2015**, *9*, 5391; b) E. Wruss, E. Zojer, O. T. Hofmann, *J. Phys. Chem. C* **2018**, *122*, 14640; c) J. Niederhausen, P. Amsalem, A. Wilke, R. Schlesinger, S. Winkler, A. Vollmer, J. P. Rabe, N. Koch, *Phys. Rev. B* **2012**, *86*, 081411; d) P. Amsalem, J. Niederhausen, A. Wilke, G. Heimel, R. Schlesinger, S. Winkler, A. Vollmer, J. P. Rabe, N. Koch, *Phys. Rev. B* **2013**, *87*, 035440.

[21] H. Li, D. He, Q. Zhou, P. Mao, J. M. Cao, L. M. Ding, J. Z. Wang, *Sci. Rep.* **2017**, *7*, 40134.

[22] R. Schlesinger, F. Bianchi, S. Blumstengel, C. Christodoulou, R. Ovsyannikov, B. Kobin, K. Moudgil, S. Barlow, S. Hecht, S. R. Marder, F. Henneberger, N. Koch, *Nat. Commun.* **2015**, *6*, 6754.

[23] a) D. C. Look, J. W. Hemsky, J. R. Sizelove, *Phys. Rev. Lett.* **1999**, *82*, 2552; b) D. Li, Y.H. Leung, A.B. Djurisic, Z.T. Liu, M.H. Xie, S.L. Shi, S.J. Xu, *Appl. Phys. Lett.* **2004**, *85*, 1601; c) D. C. Look, G. C. Farlow, Pakpoom Reunchan, Sukit Limpijumnong, S. B. Zhang, K. Nordlund, *Phys. Rev. Lett.* **2005**, *95*, 225502; d) H. B. Zeng, G. T. Duan, Y. Li, S. K. Yang, X. X. Xu, W. P. Cai, *Adv. Funct. Mater.* **2010**, *20*, 561.

[24] a) G. S. Higashi, Y. J. Chabal, G. W. Trucks, K. Raghavachari, *Appl. Phys. Lett.* **1990**, *56*, 656; b) P. Dumas, Y. J. Chabal, G. S. Higashi, *Phys. Rev. Lett.* **1990**, *65*, 1124; c) R. S. Becker, G. S. Higashi, Y. J. Chabal, A. J. Becker, *Phys. Rev. Lett.* **1990**, *65*, 1917.

[25] a) P. Hohenberg, W. Kohn, *Phys. Rev.* **1964**, *136*, B864; b) W. Kohn, L. J. Sham, *Phys. Rev.* **1965**, *140*, A1133.

[26] J. P. Perdew, K. Burke, M. Ernzerhof, *Phys. Rev. Lett.* **1996**, *77*, 3865.

[27] J. Neugebauer, M. Scheffler, *Phys. Rev. B* **1992**, *46*, 16067.

[28] a) M. Ceriotti, J. More, D. E. Manolopoulos, *Comput. Phys. Commun.* **2014**, *185*, 1019; b) V. Kapil, M. Rossi, O. Marsalek, R. Petraglia, Y. Litman, T. Spura, B. Q. Cheng, A. Cuzzocrea, R. H. Meißner, D. M. Wilkins, P. L. Juda, S. P. Bienvenue, W. Fang, J. Kessler, I. Poltavsky, S. Vandenbrande, J. Wieme, C. Corminboeuf, T. D. Kühne, D. E. Manolopoulos, T. E. Markland, J. O. Richardson, A. Tkatchenko, G. A. Tribello, V. V. Speybroeck, M. Ceriotti, *arXiv preprint* **2018**, arXiv:1808.03824.

[29] M. Ceriotti, D. E. Manolopoulos, *Phys. Rev. Lett.* **2012**, *109*, 100604.

[30] D. Nečas, P. Klapetek, *Cent. Eur. J. Phys.* **2012**, *10*, 181.

[31] R. Schlesinger, Y. Xu, O. T. Hofmann, S. Winkler, J. Frisch, J. Niederhausen, A. Vollmer, S. Blumstengel, F. Henneberger, P. Rinke, M. Scheffler, N. Koch, *Phys. Rev. B* **2013**, *87*, 155311.

[32] T. Schultz, R. Schlesinger, J. Niederhausen, F. Henneberger, S. Sadofev, S. Blumstengel, A. Vollmer, F. Bussolotti, J.-P. Yang, S. Kera, K. Parvez, N. Ueno, K. Müllen, N. Koch, *Phys. Rev. B* **2016**, *93*, 125309.

[33] W. Chen, S. Chen, D. C. Qi, X. Y. Gao, A. T. S. Wee, *J. Am. Chem. Soc.* **2007**, *129*, 10418.

[34] C. Coletti, C. Riedl, D. S. Lee, B. Krauss, L. Patthey, K. V. Klitzing, J. H. Smet, U. Starke, *Phys. Rev. B* **2010**, *81*, 235401.

[35] Y. J. Hsu, Y. L, Lai, C. H. Chen, Y. C. Lin, H. Y. Chien, J. H. Wang, T. N. Lam, Y. L. Chan, D. H. Wei, H. J. Lin, C. T. Chen, *J. Phys. Chem. Lett.* **2013**, *4*, 310.

[36] O. Rana, R. Sirvastava, G. Chauhan, M. Zulfequar, M. Husain, P. C. Srivastava, M. N. Kamalasanan, *Phys. Status Solid A* **2012**, *209*, 2539.

[37] N. Koch, S. Suhm, J. P. Rabe, A. Vollmer, R. L. Johnson, *Phys. Rev. Lett.* **2005**, *95*, 237601.

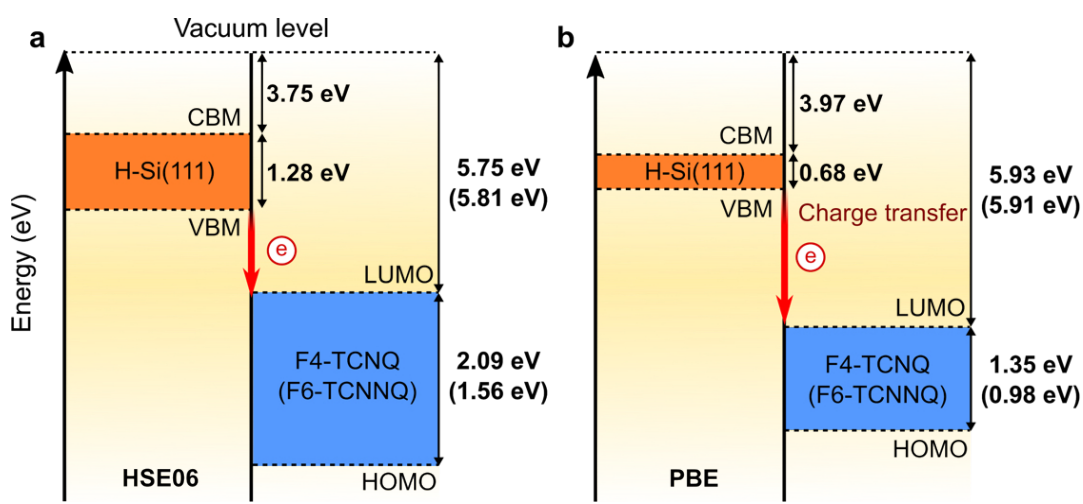

**Figure 1.** Schematic energy-level diagram for free molecules (F4-TCNQ or F6-TCNNQ) and H-Si(111), determined by different functionals, a) HSE06 and b) PBE. The band gap of H-Si(111) calculated by HSE06, 1.28 eV, is closer to the experimental value of 1.17 eV.[18]

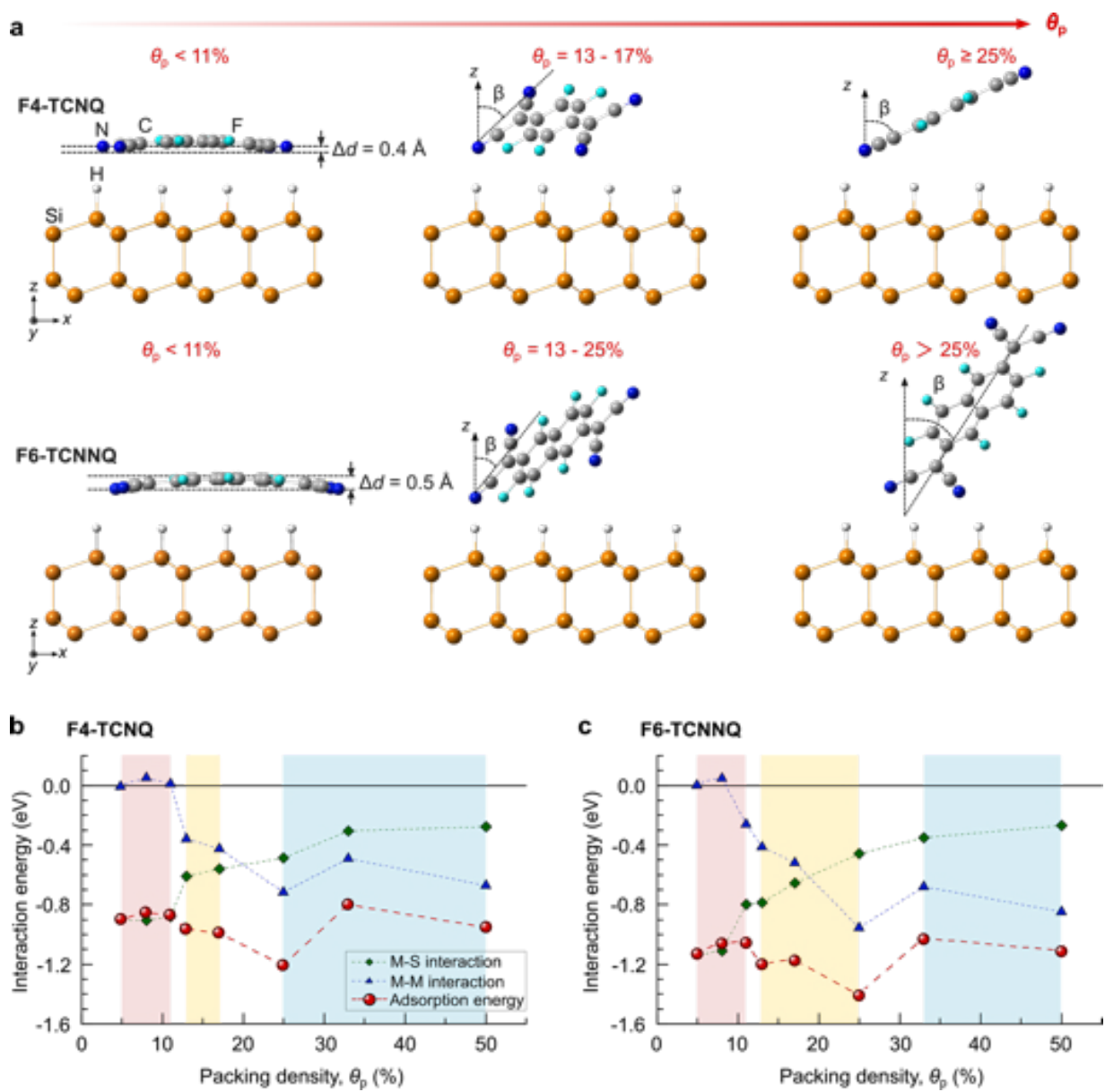

**Figure 2.** Structure and adsorption energy. a) Side views of the prevalent configurations after complete relaxation for F4-TCNQ(F6-TCNNQ)/H-Si(111) for different adsorption packing densities, $\theta_p$. $\beta$ is the tilt angle between the symmetry axis of the adsorbed molecule and the $z$-direction. The adsorption energy is marked by red spheres for b) F4-TCNQ and c) F6-TCNNQ as a function of $\theta_p$. Green squares (blue triangles) are the energies of the interaction between molecules and substrate (intermolecular interaction). Red, yellow, and green areas show the most stable lying-down, short and long-tilted configurations, respectively. The considered supercells corresponding to the different $\theta_p$ with their different adsorption phases as well as adsorption energies are listed in the Table S1 in SI.

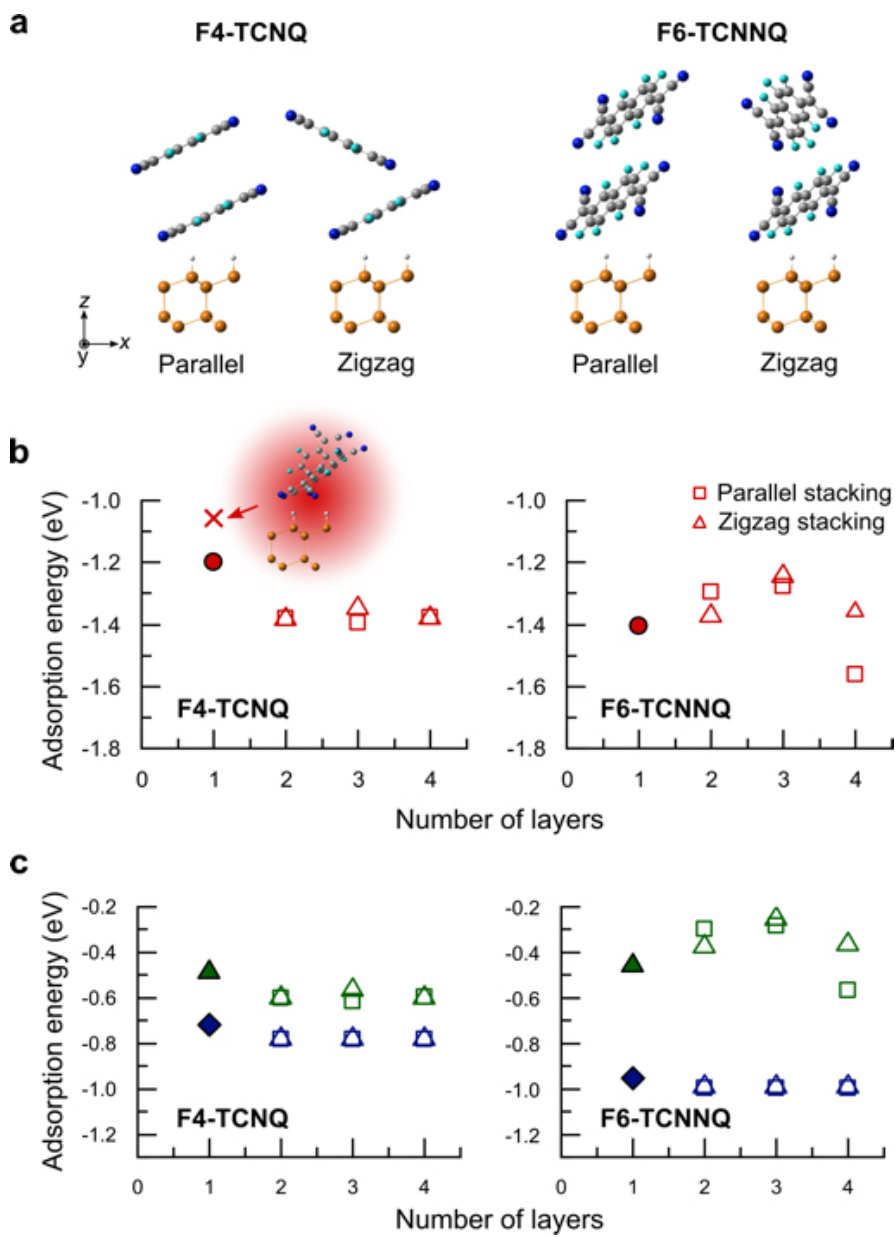

**Figure 3.** Morphologies for several adsorbed layers. a) The profile of parallel and zigzag stacking of F4-TCNQ (left panel) and F6-TCNNQ (right panel). b) The adsorption energy per molecular layer of F4-TCNQ (left) and F6-TCNNQ (right). The squares and triangles represent parallel and zigzag stacking of molecules. The cross corresponds to 33.3% packing in the first layer. c) Decomposition into interlayer (green) and intralayer (blue) of the layer adsorption energies shown in panel b).

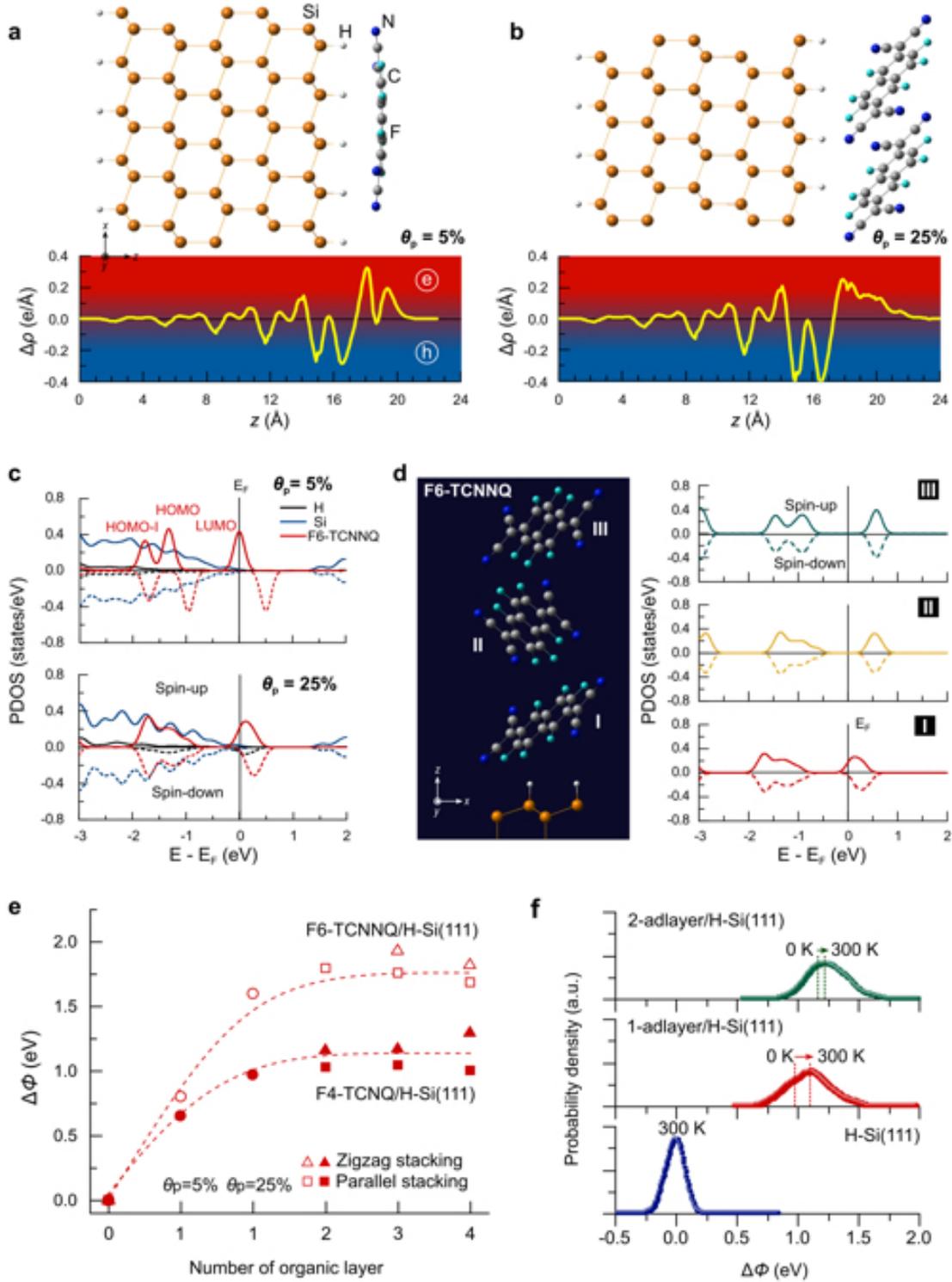

**Figure 4.** Electronic properties of F6-TCNNQ/H-Si(111) calculated with the HSE06 functional including many-body dispersion corrections. The *xy*-plane integrated charge density rearrangement, $\Delta\rho$, upon the adsorption for a) low and b) high $\theta_p$. c) Calculated PDOS at low (bottom panel) and high (upper panel) $\theta_p$. Black, blue, and red lines are the PDOS of H, Si, and molecules, respectively. d) The PDOS of each molecular layer in the zigzag model. The solid and dashed curves represent the spin-up and spin-down channels in both c) and d), and vertical lines indicate the Fermi level. e) The molecule-induced $\Delta\Phi$ for different adsorption packing density and number of layers from simulations. Circles indicate the lying-

down configuration at a packing density of 5%, and from the molecule tilted with the long-axis (F4-TCNQ) and short-axis (F6-TCNNQ) at a packing density of 25%. Triangles and squares indicate the configurations with zigzag and parallel stacking in the adsorbed multilayers, respectively. f) The probability density of ΔΦ for the bare H-Si(111) surface (blue line), one- (red) and two-adlayer of zigzag configuration (green) of F4-TCNQ on H-Si(111). Dashed lines mark the static ΔΦ at 0 K and the ΔΦ distribution at 300 K.

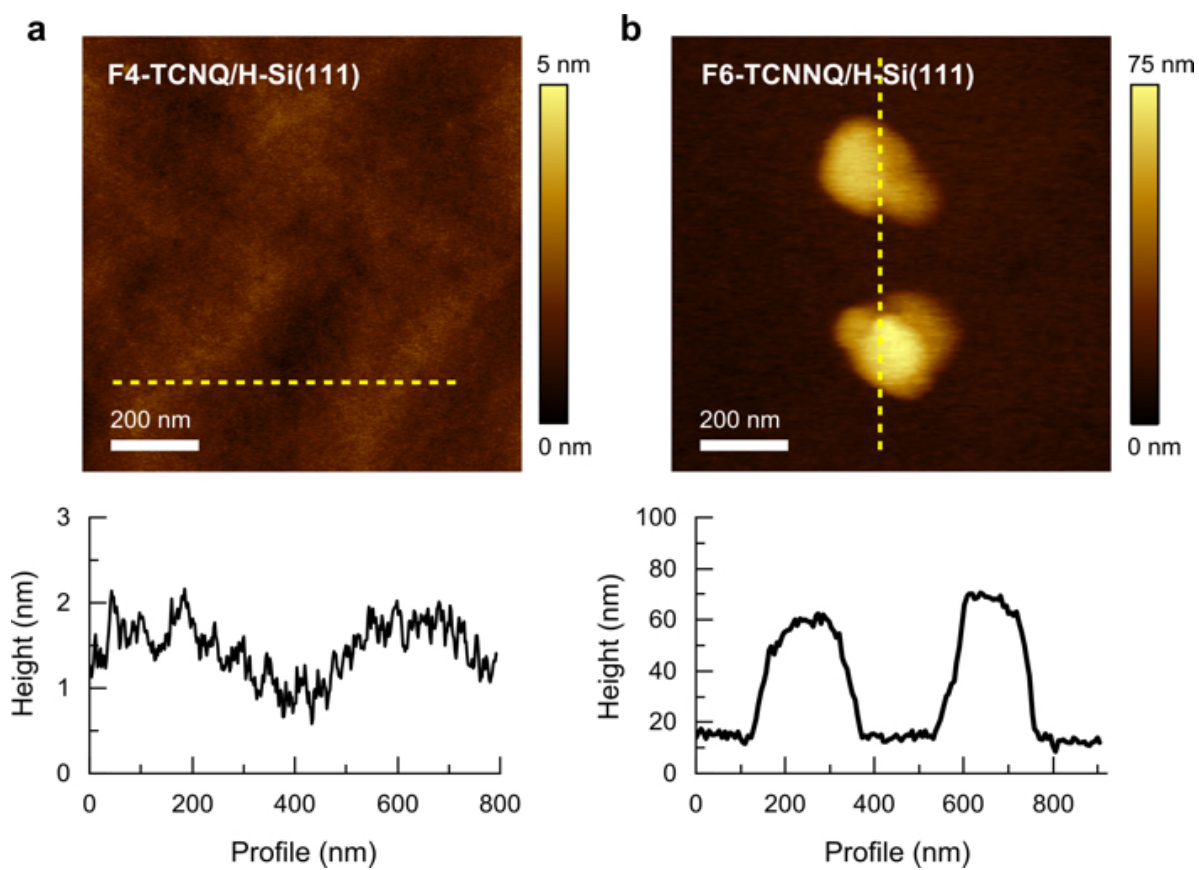

**Figure 5.** SFM images of nominally 1.2 nm F4-TCNQ and F6-TCNNQ on H-Si(111), with associated height profiles.

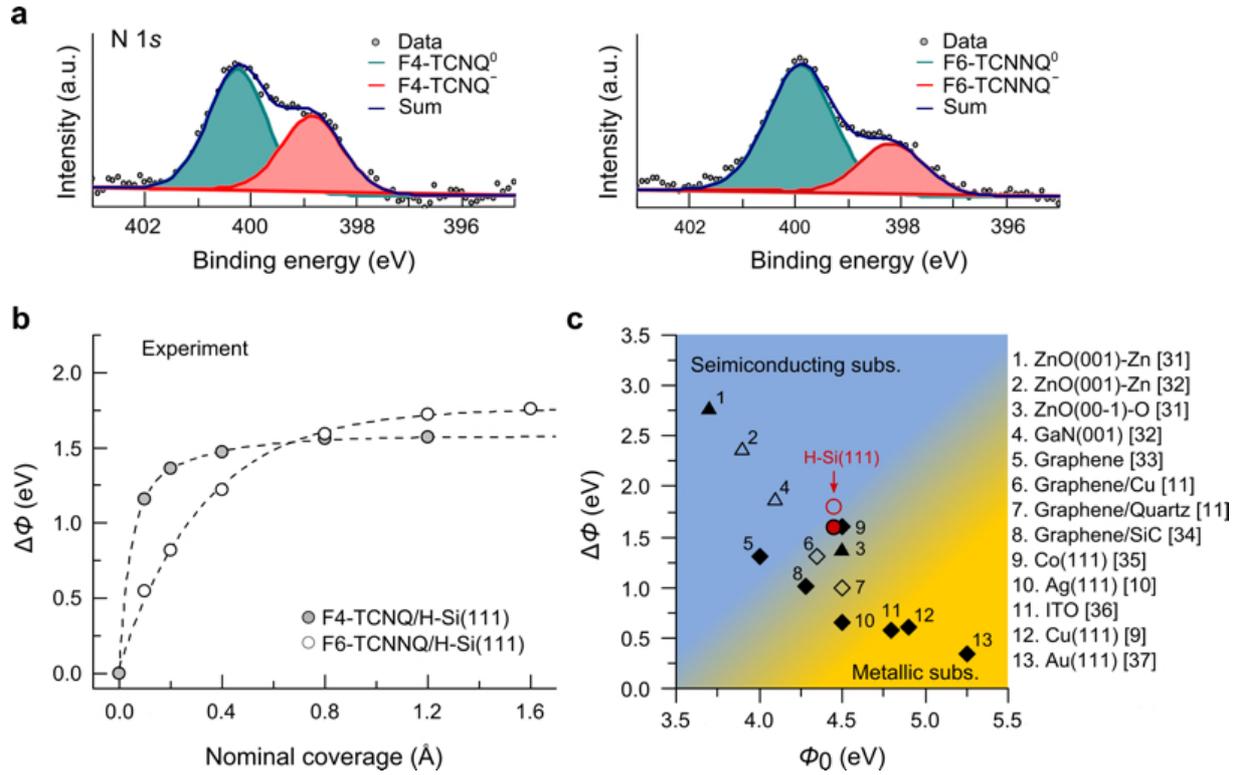

**Figure 6.** a) High resolution XPS spectra of the N 1s core level of nominally 1.2 nm of F4-TCNQ and 1.6 nm of F6-TCNNQ on H-Si(111). Two chemical states can be differentiated, the one at higher binding energy (green) stemming from neutral molecules and the one at lower binding energy (red) stemming from negatively charged molecules. b) ΔΦ at different coverages. The zero was shifted by the respective measured band-bending of the clean surface in each case. c) A comparison of initial substrate Φ and molecule-induced ΔΦ for the available data from previous experiments, with data of the present study denoted by red spheres. The triangle and diamond symbols correspond to semiconducting and metallic substrates.

**Table 1.** Work function changes (ΔΦs) and its contributions due to the molecular dipole (ΔΦ$_{MD}$) and interface charge rearrangement (ΔΦ$_{CR}$) at low and high packing density for both F4-TCNQ/H-Si(111) and F6-TCNNQ/H-Si(111) adsorbed systems.

| system | packing density ($\theta_p$) | ΔΦ (eV) | ΔΦ$_{MD}$ (eV) | ΔΦ$_{CR}$ (eV) |
|---|---|---|---|---|
| F4-TCNQ/H-Si(111) | Low - 5% | 0.65 | -0.12 | 0.83 |
| | High - 25% | 0.97 | 0 | 0.88 |
| F6-TCNNQ/H-Si(111) | Low - 5% | 0.80 | -0.15 | 0.91 |
| | High - 25% | 1.59 | 0 | 1.58 |